\documentclass{JHEP3} % 10pt is ignored!

\JHEPspecialurl{http://jhep.sissa.it/JOURNAL/JHEP3.tar.gz}

\usepackage{epsfig,multicol,bbm,amsmath}
\newcommand\fverb{\setbox\pippobox=\hbox\bgroup\verb}
\newcommand\fverbdo{\egroup\medskip\noindent%
			\fbox{\unhbox\pippobox}\ }
\newcommand\fverbit{\egroup\item[\fbox{\unhbox\pippobox}]}
\newbox\pippobox
\def\DAF{DA\char8NE}
\def\ifm#1{\relax\ifmmode#1\else$#1$\fi}
\def\ab{\ifm{\sim}}  
\makeatletter
\newdimen\z@ \z@=0pt % can be used both for 0pt and 0 
\newskip\z@skip \z@skip=0pt plus0pt minus0pt
\def\m@th{\mathsurround=\z@}
\def\ialign{\everycr{}\tabskip\z@skip\halign} % initialized \halign
\def\eqalign#1{\null\,\vcenter{\openup\jot\m@th
  \ialign{\strut\hfil$\displaystyle{##}$&$\displaystyle{{}##}$\hfil
    \crcr#1\crcr}}\,}
\makeatother
\catcode`@=11 %%At signs are letters
\newdimen\z@ \z@=0pt % can be used both for 0pt and 0
\newskip\z@skip \z@skip=0pt plus0pt minus0pt
\def\m@th{\mathsurround=\z@}
\def\ialign{\everycr{}\tabskip\z@skip\halign} % initialized \halign
\def\eqalign#1{\null\,\vcenter{\openup\jot\m@th
  \ialign{\strut\hfil$\displaystyle{##}$&$\displaystyle{{}##}$\hfil
      \crcr#1\crcr}}\,}
\catcode`@=12 % at signs are no longer letters
\def\figb#1;#2;{\parbox{#2cm}{\epsfig{file=#1.eps,width=#2cm}}}
\let\cl=\centerline
\def\figbc#1;#2;{\cl{\figb #1;#2;}}
\newcommand{\eV}{{e\kern-.07em V}}
\newcommand{\MeV}{{\rm \,M\eV}}
\newcommand{\GeV}{{\rm \,G\eV}}
\newcommand{\ns}{{\rm \,ns}}
\newcommand{\ps}{{\rm \,ps}}

\newcommand{\cm}{{\rm \,cm}}

\newcommand{\mrad}{{\rm \,mrad}}
\newcommand{\ks}  {\ensuremath{K_S}}
\newcommand{\kl}  {\ensuremath{K_L}}

\newcommand{\Fig}{Fig.\:}
\newcommand{\Eq} {Eq.\:}
\newcommand{\Se} {Sect.\:}
\newcommand{\Ta} {Table\:}
\newcommand{\phikpkm}{\mbox{$\phi\to K^+K^-$}}

\newcommand{\BR}{{\rm BR}}

\newcommand{\phif}{$\phi$~factory}
\newcommand{\Vusfo} {\ensuremath{|V_{us}\,f_+(0)|}}
\newcommand{\fo}    {\ensuremath{f_+(0)}}
\newcommand{\Vus}   {\ensuremath{V_{us}}}
\newcommand{\Vud}   {\ensuremath{V_{ud}}}

\newcommand{\g}{\ensuremath{\gamma}} 
\newcommand{\ele}[1]{\ensuremath{e^{{#1}}}}
\newcommand{\muo}[1]{\ensuremath{\mu^{{#1}}}}
\newcommand{\pai}[1]{\ensuremath{\pi^{{#1}}}}
\newcommand{\kao}[1]{\ensuremath{K^{{#1}}}}

\newcommand{\kmudue}[1]{\ensuremath{K^{{#1}}_{\mu2}}}
\newcommand{\kpidue}[1]{\ensuremath{K^{{#1}}_{\pi2}}}
\newcommand{\ketre}[1] {\ensuremath{K^{{#1}}_{e3}}}
\newcommand{\kmutre}[1]{\ensuremath{K^{{#1}}_{\mu3}}}
\newcommand{\ktaup}[1] {\ensuremath{K^{{#1}}_{3\pi}}}
\newcommand{\ketreg}[1] {\ensuremath{K^{{#1}}_{e3\gamma}}}
\newcommand{\kmutreg}[1]{\ensuremath{K^{{#1}}_{\mu3\gamma}}}
\newcommand{\kltre}[1] {\ensuremath{K^{{#1}}_{\ell 3}}}
\newcommand{\ltre}{\ensuremath{K_{\ell 3}}}
\newcommand{\Dkmudue}[1]{\ensuremath{K^{{#1}}\to\mu^{{#1}}\nu}}
\newcommand{\Dkpidue}[1]{\ensuremath{K^{{#1}}\to\pi^{{#1}}\pi^{0}}}
\newcommand{\Dketre}[1] {\ensuremath{K^{{#1}}\to\pi^0e^{{#1}}\nu}}
\newcommand{\Dketreg}[1] {\ensuremath{K^{{#1}} \to \pi^0e^{{#1}}\nu(\gamma)}}
\newcommand{\Dkmutre}[1]{\ensuremath{K^{{#1}}\to\pi^0\mu^{{#1}}\nu}}

\newcommand{\Dktaup}[1]{\ensuremath{K^{{#1}}\to\pi^{#1}\pi^0\pi^0}}

\newcommand{\Dkltreg}[1] {\ensuremath{K^{{#1}}\to\pi^0{\rm l}^{{#1}}\nu (\gamma)}}
\newcommand{\Dpimu}[1]{\ensuremath{\pi^{{#1}} \to \mu^{{#1}}\nu}}
\newcommand{\bketre}[1] {\ensuremath{{\rm BR}(K^{{#1}}_{e3})}}
\newcommand{\bkmutre}[1]{\ensuremath{{\rm BR}(K^{{#1}}_{\mu3})}}
\newcommand{\lifet}[1]{\ensuremath{{\tau}_{{\rm #1}}}}
\newcommand{\rappo}[1]{\ensuremath{\Gamma(K^{{#1}}_{\mu3})/\Gamma(K^{{#1}}_{e3})}}
\newcommand{\rme}[1]{\ensuremath{R^{#1}_{\mu e}}}
\newcommand{\pb}     {\ensuremath{{\rm pb^{-1}}}}
\newcommand{\mdue}   {\ensuremath{m^{2}_{{\rm lept}}}}
\newcommand{\mtrenta} {\ensuremath{p^{\ast}_{\pi\mu}}}
\newcommand{\pstarp} {\ensuremath{p^{\ast}_{\pi}}}
\newcommand{\pstarm} {\ensuremath{p^{\ast}_{\mu}}}
\newcommand{\emiss} {\ensuremath{E_{{\rm miss}}-p_{{\rm miss}}}}
\newcommand{\plab} {\ensuremath{p_{\rm lab}}}
\newcommand{\eff}[1]{\ensuremath{\epsilon_{#1}}}
\newcommand{\effi}[2]{\ensuremath{\epsilon_{{\rm #1}}^{{\rm #2}}}}
\newcommand{\cdue}{\ensuremath{\chi^2}}
\newcommand{\cveto} {CV} 
\newcommand{\filfo} {FilFo} 
\newcommand{\atb} {\ensuremath{\alpha_{{\rm TB}}}}
\newcommand{\lep} {\ensuremath{_{{\rm lept}}}}
\newcommand{\rxy} {\ensuremath{\rho_{{\rm vtx}}}}
\newcommand{\tk} {\ensuremath{\theta_K}}
\newcommand{\tope} {\ensuremath{\omega}}
\newcommand{\rcorr}[1] {\ensuremath{r_{{\rm #1}}}}
\newcommand{\pk} {\ensuremath{p_{_{{\rm k}}}}}
\newcommand{\ntkmudue}{\ensuremath{K^{}_{\mu2,0}}}
\title{\mathversion{bold}Measurement of the absolute branching ratios for 
semileptonic $K^\pm$ decays with the KLOE detector}
\author{The KLOE collaboration:\\
F.~Ambrosino,$^d$
A.~Antonelli,$^a$
M.~Antonelli,$^a$
F.~Archilli,$^a$
C.~Bacci,$^g$
P.~Beltrame,$^b$
G.~Bencivenni,$^a$
S.~Bertolucci,$^a$
C.~Bini,$^f$
C.~Bloise,$^a$
S.~Bocchetta,$^g$
F.~Bossi,$^a$
P.~Branchini,$^g$
R.~Caloi,$^f$
P.~Campana,$^a$
G.~Capon,$^a$
T.~Capussela,$^a$
F.~Ceradini,$^g$
S.~Chi,$^a$
G.~Chiefari,$^d$
P.~Ciambrone,$^a$
E.~De~Lucia,$^a$
A.~De~Santis,$^f$
P.~De~Simone,$^a$
G.~De~Zorzi,$^f$
A.~Denig,$^b$
A.~Di~Domenico,$^f$
C.~Di~Donato,$^d$
B.~Di~Micco,$^g$
A.~Doria,$^d$
M.~Dreucci,$^a$
G.~Felici,$^a$
A.~Ferrari,$^a$
M.~L.~Ferrer,$^a$
S.~Fiore,$^f$
C.~Forti,$^a$
P.~Franzini,$^f$
C.~Gatti,$^a$
P.~Gauzzi,$^f$
S.~Giovannella,$^a$
E.~Gorini,$^c$
E.~Graziani,$^g$
W.~Kluge,$^b$
V.~Kulikov,$^j$
F.~Lacava,$^f$
G.~Lanfranchi,$^a$
J.~Lee-Franzini,$^{a,h}$
D.~Leone,$^b$
M.~Martini,$^a$
P.~Massarotti,$^d$
W.~Mei,$^a$
S.~Meola,$^d$
S.~Miscetti,$^a$
M.~Moulson,$^a$
S.~M\"uller,$^a$
F.~Murtas,$^a$
M.~Napolitano,$^d$
F.~Nguyen,$^g$
M.~Palutan,$^a$
E.~Pasqualucci,$^f$
A.~Passeri,$^g$
V.~Patera,$^{a,e}$
F.~Perfetto,$^d$
M.~Primavera,$^c$
P.~Santangelo,$^a$
G.~Saracino,$^d$
B.~Sciascia,$^a$
A.~Sciubba,$^{a,e}$
A.~Sibidanov,$^{a}$
T.~Spadaro,$^a$
M.~Testa,$^f$
L.~Tortora,$^g$
P.~Valente,$^f$
G.~Venanzoni,$^a$
R.Versaci,$^a$
G.~Xu,$^{a,i}$
\\
\llap{$^a$}Laboratori Nazionali di Frascati dell'INFN, Frascati, Italy\\
\llap{$^b$}Institut f\"ur Experimentelle Kernphysik, Universit\"at Karlsruhe, Germany\\
\llap{$^c$}Dipartimento di Fisica dell'Universit\`a e Sezione INFN, Lecce, Italy\\%%%                               h=>f
\llap{$^d$}Dipartimento di Scienze Fisiche dell'Universit\`a  ``Federico II'' e Sezione INFN, Napoli, Italy\\
\llap{$^e$}Dipartimento di Energetica dell'Universit\`a ``La Sapienza'', Roma, Italy\\%%%                       m=j
\llap{$^f$}Dipartimento di Fisica dell'Universit\`a ``La Sapienza'' e Sezione INFN, Roma, Italy\\
\llap{$^g$}Dipartimento di Fisica dell'Universit\`a ``Roma Tre'' e Sezione INFN, Roma, Italy\\
\llap{$^h$}Physics Department, State University of New York at Stony Brook, USA\\%%%                        l=>i
\llap{$^i$}Institute of High Energy Physics of Academia Sinica,  Beijing, China\\%%%                                j=>h
\llap{$^j$}Institute for Theoretical and Experimental Physics, Moscow, Russia\\%%%                                  i=>g
}
\received{\today} 		%%
\revised{\today}
\accepted{\today}		%% These are for published papers.

\abstract{
Using a sample of over 600 million \phikpkm\ decays collected at the \DAF\ \ele{+}\ele{-} collider,
we have measured with the KLOE detector the absolute branching ratios for the charged kaon semileptonic decays,
\Dketre{\pm}(\g)  and \Dkmutre{\pm}(\g).
The results, $\bketre{} = 0.04965 \pm 0.00038_{\rm stat}\pm 0.00037_{\rm syst}$ 
and $\bkmutre{} = 0.03233 \pm 0.00029_{\rm stat}\pm 0.00026_{\rm syst}$, are inclusive of radiation.
Accounting for correlations, we derive the ratio $\rappo{} = 0.6511\pm0.0064$.
Using the semileptonic form factors measured in the same experiment, we obtain \Vusfo\ = $0.2141 \pm 0.0013$.}

\keywords{e$+$e$-$ experiments}

\begin{document}
\section{Introduction}

At present, the determinations of \Vus\ and \Vud\ provide the most precise verification 
of the unitarity of the CKM matrix.
The relation $1 - |V_{ud}|^2 - |V_{us}|^2  - |V_{ub}|^2  = 0$ can be tested 
with an absolute accuracy of few parts per mil using $|V_{ud}|$ as measured in nuclear beta 
decays and $|V_{us}|$ as derived from semileptonic kaon decays.
Since it was already known in 1983 that $|V_{ub}|^2<4\times10^{-5}$~\cite{Vub} 
and today $|V_{ub}|^2$ is $\sim1.5\times 10^{-5}$~\cite{PDG06},
$|V_{ub}|^2$ will be ignored in the following.
All experimental inputs to \Vus\ --- branching ratios (BRs), lifetimes, and form 
factors --- can be measured with the KLOE detector.
Using tagging techniques, we have already measured the complete set of 
inputs for \kl\ decays~\cite{KLOEklbr,KLOEkllife,KLOEklff,KLOEklffmu}, 
\BR(\ketre{}) for the \ks~\cite{KLOEkse3}, and the absolute BRs 
for \Dkmudue{\pm}~\cite{KLOEkmudue} and 
\Dktaup{\pm}~\cite{KLOEktaup} decays.
Here, we report on the measurement of the absolute BRs for the decays 
\Dketre{\pm}(\g) (\ketre{}) and \Dkmutre{\pm}(\g) (\kmutre{}).
Our measurements, which make use of a tagging technique,
are fully inclusive of final-state radiation.

\section{Experimental setup}
\label{sec:KLOE}

The data were collected with KLOE detector at \DAF, the Frascati \phif. \DAF\ is an \ele{+}\ele{-} collider
which operates at a center of mass energy of \ab1020~\MeV, the mass of the $\phi$ meson.
Positron and electron beams of equal energy collide at an angle of ($\pi-25~\mrad$), producing $\phi$ mesons with a small 
momentum in the horizontal plane, $p_\phi \ab13$~\MeV.
$\phi$ mesons decay \ab49\% of the time into nearly collinear \kao{+}\kao{-} pairs; the detection of a \kao{\mp} meson (the tagging kaon) therefore signals the presence of a \kao{\pm} (the tagged kaon) 
independently of its decay mode.
This technique is called \kao{\pm} tagging in the following.
The results presented here are based on an integrated luminosity of about 410~\pb\ delivered by \DAF\ in 2001-02,
corresponding to \ab$6\times10^8$ \kao{+}\kao{-} pairs produced.

The KLOE detector consists of a large cylindrical drift chamber surrounded by a lead scintillating-fiber electromagnetic calorimeter.
A superconducting coil around the calorimeter provides a 0.52 T field. 
The drift chamber (DC) \cite{DC} is 4~m in diameter and 3.3~m long.
The momentum resolution for tracks at large polar angles is $\sigma_{p_{\perp}}/p_{\perp}\approx 0.4\%$.
The vertex between two intersecting tracks is reconstructed with a spatial resolution of \ab3~mm.
The calorimeter (EMC) \cite{EMC} is divided into a barrel and two endcaps. 
It is segmented in depth into five layers and covers 98\% of the solid angle.
Energy deposits nearby in time and space are grouped into calorimeter clusters. The energy and time resolutions are $\sigma_E/E = 5.7\%/\sqrt{E\ (\GeV)}$ and
$\sigma_t = 57 \ps/\sqrt{E\ (\GeV)}\oplus100 \ps$, respectively.
The trigger~\cite{TRG} uses only calorimeter information.
Two energy deposits above threshold ($E>50$ \MeV\ for the barrel and  
$E>150$ \MeV\ for endcaps) are required.
Recognition and rejection of cosmic-ray events is also performed at the trigger level. Events with two energy
deposits above a 30 \MeV\ threshold in the outermost calorimeter plane 
are rejected.

To reject residual cosmic rays and machine background events, we use
an offline software filter (\filfo) that exploits calorimeter information before tracks are reconstructed. 
As an example, the filter tests the hypothesis that time difference between pair of clusters be 
compatible with the time of flight of a muon crossing the detector.
The response of the detector to the decays of interest and the various 
background sources  
were studied by using the KLOE Monte Carlo (MC) simulation program~\cite{OfflineNIM}.
Changes in machine parameters and background conditions are simulated on a run-by-run basis 
in order to properly track the frequent changes in machine operation.
The MC sample of \phikpkm\ decays used for the present analysis 
corresponds to an integrated luminosity of about 480~\pb;
the sample for the other $\phi-$meson final states is equivalent in statistics
to \ab90~\pb\ of integrated luminosity.

\section{Method of measurement}
\label{sec:meto}

The use of a tagging technique allows the 
measurement of absolute branching ratios.
Reconstruction of one of the two-body decays 
\Dkmudue{\mp} (\kmudue{}) and \Dkpidue{\mp} (\kpidue{}) in an event
signals the presence of a \kao{\pm}; this provides a clean, counted
sample of \kao{\pm} decays
from which to select signal events (\ketre{\pm} or \kmutre{\pm} decays).
Let $N_{\ltre}$ be the number of events identified as \ketre{} or \kmutre{}
in a given tagged sample,
and $N_{{\rm tag}}$ the total number of tagged \kao{\pm} events in the sample. 
The branching ratio of each signal decay, \ketre{} or \kmutre{}, can be determined as:
\begin{equation}
  \label{eq:master}
  BR(\kltre{}) = \frac{N_{\ltre}}{N_{{\rm tag}}\eff{\ltre}} \atb {\mbox ,}
\end{equation}
where \eff{\ltre} is the identification efficiency for semileptonic decays, given the tag.
This efficiency includes the detector acceptance (\effi{FV}{}),
and the reconstruction and selection efficiencies for \kltre{} events (\effi{Sel}{}). 

\effi{FV}{} is corrected for losses of \kao{\pm} from nuclear interactions in the material traversed 
by kaons before entering the DC.
This material includes the beam pipe 
(50~$\mu$m of Be and 500~$\mu$m of \textit{AlBe-met}, an alloy of 40\% of Be and 60\% of Al) 
and the inner DC wall
(750 $\mu$m of C and 200 $\mu$m of Al).
The probability of interaction in the KLOE setup is negligible (\ab$10^{-5}$) for \kao{+}, while it is 
\ab3.4\% for \kao{-}, as estimated by MC. Therefore, this correction is necessary only for samples tagged by \kao{+} decays.

The quantity \atb, which we refer to as the tag bias in the following, accounts for the slight dependence of 
the reconstruction and identification efficiency
for the tagging \kmudue{} (or \kpidue{}) decay on the decay mode of the 
tagged kaon. 
Both \eff{\ltre} and \atb, are evaluated from MC, with corrections evaluated from data and MC control samples.

\section{Tag selection}
\label{sec:tag}

In the $\phi$ center of mass, the two kaons are produced back-to-back with 
momentum $\pk\ab127$~\MeV.
Since the $\phi$ has a transverse momentum of 13~\MeV,
in the laboratory frame \pk\ ranges between 120~\MeV\ and 133~\MeV.
The \kao{\pm} decay length $\lambda_\pm$ is \ab 95~\cm.
Before entering the DC, kaons have to pass through the beam pipe and 
through the DC inner wall, and lose about 5~\MeV\ of energy. 
As a result, for kaons in the DC \pk\ is about 100~\MeV, and the 
decay length is reduced to about 75~\cm.

Two-body decays are observed as vertices in the drift chamber between an incoming track (the kaon) 
and an outgoing track of the same charge.
Kaons are identified as tracks with momentum 70$<$\pk$<$130~\MeV\
whose point of closest approach to the \ele{+}\ele{-} collision point (IP)
lies inside a cylinder 10~\cm\ in radius and 20~\cm\ in length along 
the $z$-axis.\footnote{$x$ and $y$ are the coordinates on the plane perpendicular to the beam axis; $z$ 
is the coordinate along the beam axis.}
The kaon decay vertex must be reconstructed within a fiducial volume (FV) defined as 
a cylinder of radius 40$<$\rcorr{xy}$<$150 \cm\ and length $|z|<$130~\cm, centered on the IP and
coaxial with the beams.
A kaon has a probability $\effi{FV}{}\simeq0.56$ of decaying in the FV as determined by MC.
The combined reconstruction efficiency for the kaon and secondary tracks connected with a vertex 
(which we refer to as the decay chain in the following) is about 0.6 as estimated by MC.

The momentum of the secondary track is computed in the kaon rest frame
using the muon and pion mass hypotheses, \pstarm\ and \pstarp, respectively.
The resulting momentum distributions are shown in \Fig\ref{fig:pstar_mupi}.
\kmudue{} events are identified as having 231$<$\pstarm$<$241~\MeV\ 
(the shaded area of \Fig\ref{fig:pstar_mupi}, left), while
\kpidue{} event candidates are identified as having 201$<$\pstarp$<$209~\MeV\
(as in \Fig\ref{fig:pstar_mupi}, right).
The tails in the distributions are due to resolution effects, and some 
residual semileptonic contamination on the left of the \kpidue{} peak.
\FIGURE{\figb pstarkm2_lin;7.;\kern.4cm\figb pstarkp2_lin;7.;
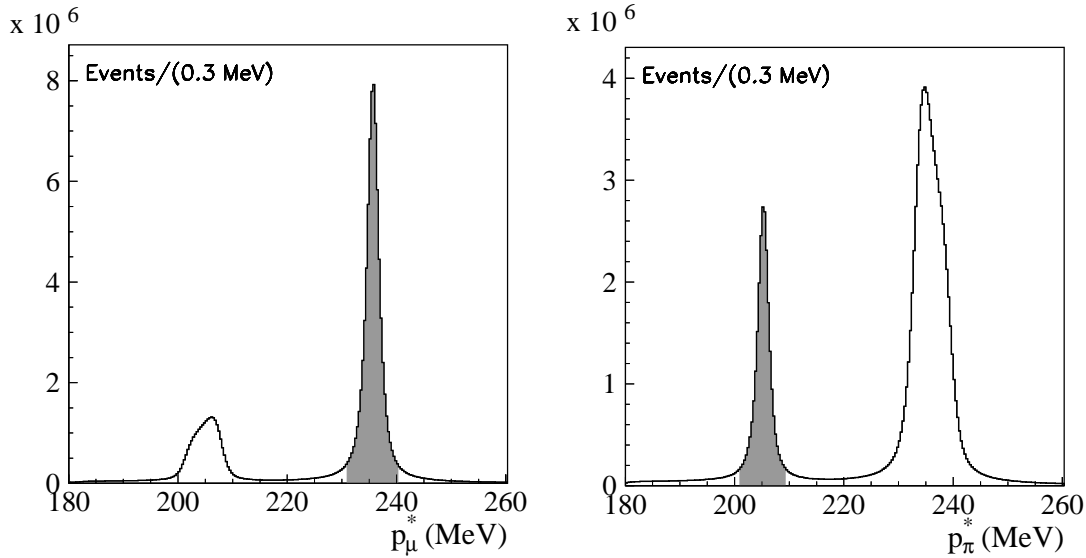
  \caption{Momentum distribution of the secondary track in the kaon rest
    frame using the muon (left) and pion (right) mass. The shaded peaks correspond to events selected
    as \kmudue{} (left) and \kpidue{} (right) decays.\label{fig:pstar_mupi}}}
The secondary track is extrapolated to the calorimeter surface and associated to a calorimeter cluster, if possible.
%MATT
The association is based on the distance between the impact point of the track on the calorimeter and the nearest cluster;
a cut is made on the component of this distance in the plane orthogonal to the direction of incidence of the track.
The efficiency and the acceptance for the extrapolation, together with the efficiency for the association, 
is about 0.7 as estimated by MC. 

To reduce the tag bias, 
we require the tagging decay to satisfy the calorimeter trigger
by itself. 
\kmudue{} decays can independently generate a trigger when the muon
incident on the calorimeter traverses two nearby trigger sectors. 
This happens in about 30\% of events with identified \kmudue{} decays.
For the remaining events identified as containing a \kmudue{} decay (referred to as 
the \ntkmudue\ sample in the following) the tag bias correction is
large ($1 - \atb \ab 0.10$). We use these events only as a control sample. 
For \kpidue{} events, the calorimeter trigger can be satisfied by the
two photon clusters from the \pai{0}. To identify \kpidue{} events, 
we require the \pai{0} to be reconstructed as follows.
For each cluster with E$>$50~\MeV\ not associated to any track,
the kaon decay time $t^K_{\gamma,i}$ is calculated using the 
cluster time $t_{cl}^i$ and the distance $L_i$ between the 
\kao{\pm} decay vertex and the cluster position:
%Old%$\delta t^K_{\gamma,i} = t_{cl}^i - L_i / c$.  
$t^K_{\gamma,i} = t_{cl}^i - L_i / c$.  
%Old%This time difference should have the same value for two photons 
This time should have the same value for two photons 
from the same \pai{0} decay, so we require the presence of two clusters
%Old%for which $|\delta t^K_{\gamma,1}-\delta t^K_{\gamma,2}|<3\sigma_t$. 
for which $|t^K_{\gamma,1}-t^K_{\gamma,2}|<3\sigma_t$ (see \Se\ref{sec:KLOE}). 
Using the energies and the positions of the two clusters,
the \g\g\ invariant mass is calculated and a 3$\sigma$ cut ($\sigma$\ab18\MeV)
about the nominal value of the \pai{0} mass is used to identify the \pai{0} 
from a \kpidue{} decay.
The calorimeter trigger is satisfied if the two identified photons 
fire two different trigger sectors.
The combined probability for a \kpidue{} decay to be identified and to 
independently satisfy the trigger is about 0.25 as determined by MC.

The overall efficiency for the identification of the tagging kaon ranges between \ab4.4\% and \ab5.7\% depending on the sample.
In the data set analyzed, about 60 million tagging decays were identified 
and divided into the four independent tag samples listed in \Ta\ref{tab:ntag}.
\TABLE{
  \begin{tabular}{c||cccc}\hline
    Tag sample       & \kmudue{+}     & \kpidue{+}    & \kmudue{-}     & \kpidue{-}     \\ \hline
    N$_{{\rm tag}}$  & $21\;319\;804$ & $7\;220\;354$ & $21\;874\;232$ & $6\;904\;949$  \\ \hline
  \end{tabular}
  \caption{Number of events selected for each tag type.}
  \label{tab:ntag}}
MC studies show that the contamination due to $\phi$ decays other than \kao{+}\kao{-} is negligible.

\section{Tag bias}
\label{ssec:tb}

Ideally, the efficiency for the identification of a tagging kaon would 
not depend on the decay mode of the tagged kaon. 
In reality, however, the geometrical overlap of the ``tag'' and ``signal'' 
parts of the \kao{+}\kao{-}
event and the fact that the trigger, offline background filter (\filfo), and
tracking procedures look at the event globally, 
make the separation into two distinct topologies arbitrary. 
The \kao{\pm}\ tagging efficiency is not completely independent of 
the \kao{\mp}\ decay mode, and the tag bias must be precisely determined.
The factor \atb\ in \Eq\ref{eq:master} is defined as
\begin{equation}
  \label{eq:tb}
 \atb = \frac{\sum_{i}{f(i)\effi{tag}{}(i)}}{\effi{tag}{}(\kltre{})}{\mbox ,}
\end{equation}
where \effi{tag}{}($i$) is the tagging efficiency given that the tagged kaon 
evolves to a final state $i$.
In the sum, $i$ indexes all possible outcomes $i$ occurring with probability 
$f(i)$ for the signal kaon, including not only all decay modes, but also
possibly nuclear interactions with the beam pipe or inner DC wall.
If the efficiency \effi{tag}{}($i$) were the same for all $i$, \atb\ 
would be equal to unity.
As noted in \Se\ref{sec:tag}, one of the main sources of tag bias is 
the dependence of the trigger efficiency on the decay mode of the tagged kaon;
the requirement that the tagging kaon independently satisfy
the trigger makes \effi{trg}{}=1, decreasing the tag bias.
\atb\ refers to the tag bias from other sources, and can be estimated 
only by using the MC. The values of \atb\ for each combination of 
tag and signal decay mode are listed in \Ta\ref{tab:tb}.
\TABLE{
  \begin{tabular}[c]{c||cccc}\hline
                & \kmudue{+}     & \kpidue{+}     & \kmudue{-}     & \kpidue{-}     \\ \hline
    \ketre{}    & 0.9694(1)(5) & 1.0137(3)(5) & 0.9884(1)(5) & 1.0328(2)(3) \\ 
    \kmutre{}   & 0.9756(1)(5) & 1.0210(4)(5) & 0.9963(1)(5) & 1.0371(2)(3) \\ \hline 
  \end{tabular}
  \caption{\atb\ computed by MC and corrected for data-MC differences.
    The statistical and systematic errors on the last digit are shown 
    in parentheses.}
  \label{tab:tb}}
The values of \atb\ range from about 0.97 to 1.04, depending on the tag 
sample used, and include a small correction due to differences in the 
performance
of the cosmic-ray veto and offline background filter in data and in MC.
The determination of the systematic errors is discussed in \Se\ref{sec:sys}.

\section{\mathversion{bold}Search for semileptonic \kao{\pm} decays}
\label{sec:kl3}

For the selection of signal events, we require the reconstruction of 
the vertex between the kaon and secondary tracks in the DC, and 
of two clusters from a \pai{0} originating at this vertex.
The criteria are the same as those used for identification of the tagging
decay.
The average efficiency for complete reconstruction of the decay chain
is \ab60\% as evaluated by MC.
This estimate is corrected for differences between data and MC 
in the tracking efficiency using data and MC samples of 
\kao{\pm}$\to$\pai{0}$X$ events
as described in \Se\ref{sec:sys}.
The average correction factor applied to the MC efficiency is \ab0.87.

The secondary track is extrapolated to the calorimeter and geometrically 
associated to a cluster.
The efficiency for the association is more than 99\% for \ketre{} 
events and about 91\% for 
\kmutre{} events. 
Both have been estimated using MC. The correction due to data-MC
discrepancies is
0.4\% for \ketre{}
(measured using a sample of \kl$\to\pi e \nu$ events, as described in~\cite{KLOEklff}), 
and 2\% for \kmutre{}.
The latter value is obtained by  
combining the corrections measured using samples of 
\kl$\to\pi\mu\nu$ (see~\cite{KLOEklffmu}) and \kmudue{\pm} events.

The fiducial volume efficiency, \effi{{\rm FV}}{}, is about 56\%, as in the tag selection,
and, for $K^-$'s, is corrected for nuclear interactions.
This correction has been checked using data, because of lack of
knowledge of \kao{\pm}-nuclear interaction cross sections for \pk$<$1~\GeV. 
Since the geometrical efficiency for the detection of \kao{\pm} decays depends 
on the \kao{\pm} lifetime \lifet, so do the values of the BRs:
$\BR(\tau)/\BR(\tau^{(0)}) = 1 - 0.0364\,\ns^{-1}\:\left(\tau - \tau^{(0)}\right)$,
where $\tau^{(0)}= 12.385\pm0.024$~\ns,
the current world average value~\cite{PDG06}.

To reject the abundant two body decays, we require $\pstarp<192$~\MeV\
for signal events.
Only poorly reconstructed \kmudue{} and \kpidue{} decays and \kpidue{} events 
with an early \Dpimu{\pm} decay survive this cut. 
The procedure used to identify the \pai{0} associated to the decay vertex
is similar to that used in the selection of \kpidue{} decays,
the only difference being that for signal events we require
$E > 20$~\MeV\ for each cluster.
The efficiency for \pai{0} identification 
(including EMC acceptance and cluster efficiency) 
is about 0.57 as estimated by MC. 
The single-photon detection efficiencies for data and MC are evaluated as a 
function of photon energy
using \kpidue{\pm} events; their ratio is used to correct the MC efficiency.
The average correction factor is \ab0.98.

After \pai{0} selection,
the sample is composed mainly of semileptonic decays, with residual
contamination from \kpidue{} and \Dktaup{\pm} (\ktaup{}) decays.
To reject \kpidue{} events in which the \pai{\pm} decays to
$\muo{\pm}\nu$ before entering the DC,
we evaluate the lepton momentum using the $m_\mu$ mass hypothesis (\mtrenta) in the center of mass of the \pai{\pm}.
The \pai{\pm} momentum is defined as the missing momentum at the decay vertex, \pk$-p_{\pai{0}}$.
By requiring $\mtrenta > 60$~\MeV, we reject about 95\% of 
\pai{\pm}$\to$\muo{\pm}$\nu$ decays while retaining
about 83\% of \ketre{} and 78\% of \kmutre{} events, 
as estimated by MC.
The contamination from \ktaup{} events is reduced by requiring \emiss,
calculated using the $m_e$ mass hypothesis, to be less than 90~\MeV.
After the above cuts, the contamination from non-\kltre{} events is about 2.1\% in each tag sample,
and consists of
\ab1.4\% \kpidue{} decays and \ab0.7\% \ktaup{} decays;
for \kao{-}, a contamination of \ab0.3\% from nuclear interactions is also
present.

\TABLE{
  \begin{tabular}[c]{c||c||c}\hline
              & \ketre{} & \kmutre{}\\ \hline
   \kmudue{+} & 0.0957(11)(6) & 0.0815(13)(4)\\       
   \kpidue{+} & 0.0989(16)(6) & 0.0848(19)(4)\\       
   \kmudue{-} & 0.0983(12)(6) & 0.0841(9)(4)\\        
   \kpidue{-} & 0.1008(18)(6) & 0.0867(20)(4)\\ \hline
  \end{tabular}
  \caption{\eff{\ltre} efficiency value corrected for data and MC differences. 
    Statistical and systematic errors are shown in parentheses.}
  \label{tab:efficorr_erro}}
The overall efficiencies for reconstruction and identification of \kltre{}
events (\eff{\ltre} in \Eq\ref{eq:master}), including data-MC corrections,
are listed by decay and tag type in \Ta\ref{tab:efficorr_erro}.
The statistical errors account for both the MC statistics and 
the statistics of the control samples used to estimate the 
data-MC efficiency corrections.
The uncertainties from control sample statistics represent the largest contributions (about 1\%) 
to the total errors on the \BR\ measurements.
In particular, for \ketre{} the dominant uncertainty is from the tracking 
correction, while for \kmutre{}, the uncertainties from the tracking and 
muon cluster corrections are at the same level.
Further details are given in \Se\ref{sec:sys}.

To isolate \ketre{} and \kmutre{} decays, the lepton is identified using a 
time-of-flight technique. 
Specifically, if the secondary track is given the correct mass assignment, 
the kaon decay time estimated using the cluster associated to this track
($t\lep^K$)
must be equal to the decay time estimated from the photon clusters from the
\pai{0} ($t_{\pai{0}}^K$).
We calculate $t\lep^K$ as $t\lep - L\lep/\beta\lep c$, where
$t\lep$ is the arrival time of the cluster associated to the secondary track 
and $\beta\lep$ and $L\lep$ are the velocity and length for this track, 
respectively.
The lepton mass is then obtained by imposing $t_{\pai{0}}^K = t\lep^K$:
$$
  \mdue ~=~ p\lep^2 
  \left [ \frac{c^2}{L\lep^2}  \left (t\lep - t_{\pai{0}}^K \right )^2 - 1 \right ] {\mbox ,}
$$
where $p\lep$ is the momentum in the laboratory frame.
The \mdue\ distribution is shown in \Fig\ref{fig:ke3peak} left for signal and background MC events.
The \ketre{} and \kmutre{} signals are evident. 
There is a residual background of \ab2\%; this is not visible in the 
distribution and is shown in the inset.
\kpidue{} and \ktaup{} decays contribute to the broad background 
distribution. For \kao{-}'s, nuclear interactions also contribute.
The small peak at \mdue\ equal zero is due to the incorret association of the track with a background cluster.

\FIGURE{\figb m2compo_t28lin_inset;7.;\kern.4cm\figb 2007_kl3_tag28_meto1_compolog;7.;
  \caption{Left: MC \mdue\ distribution for \kmudue{+}-tagged semileptonic and background events; 
    the \ab2\% residual background is not visible and is shown in the inset.
    Right: \mdue\ distribution in logarithmic scale for \kltre{} candidate 
    events identified in the \kmudue{+}-tagged sample
    for data (triangles) and MC after fit (shaded histograms).\label{fig:ke3peak}}}
\TABLE{
  \begin{tabular}[c]{c||cccc}\hline
                   & \kmudue{+}      & \kpidue{+}      & \kmudue{-}      & \kpidue{-} \\ \hline
    N$_{Ke3}$      & 101$\;$733 (411) & 34$\;$109 (243) & 108$\;$125 (430) & 33$\;$887 (243)  \\ 
    N$_{K\mu3}$    &  55$\;$919 (339) & 18$\;$999 (200) &  59$\;$730 (358) & 18$\;$923 (205)  \\ \hline 
  \end{tabular}
  \caption{Summary of fit results for the observed numbers of \kltre{} events.
    Errors are obtained from the fit and account for data and MC statistics.} 
  \label{tab:yield}}
The numbers of signal events are obtained from a fit to the \mdue\ data spectrum 
with a linear combination of the MC distributions for \ketre{} decays,
\kmutre{} decays, and background.
The fit parameters are the numbers of signal and background events. 
The result of the fit for the \kmudue{+}-tagged sample
is shown in figure~\ref{fig:ke3peak}, right. 
The results for each of the four tag samples are
summarized in \Ta\ref{tab:yield}.
For all samples, the fit gives a \ab1\% correlation 
between the numbers of \ketre{} and \kmutre{} events.

\FIGURE{\parbox{7cm}{\figbc ke3gamma_relateff_spe;6.8;}
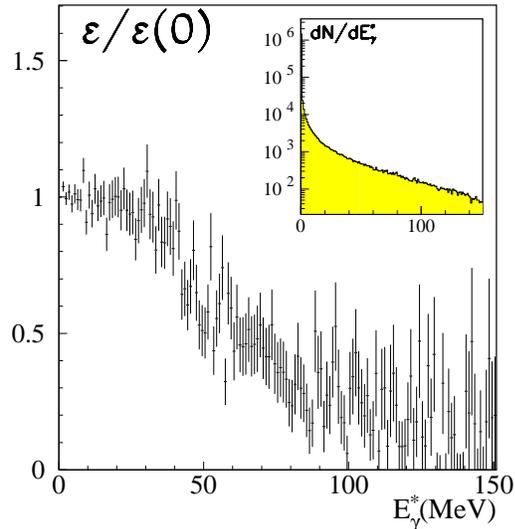
  \caption{\eff{\ltre} for \Dketreg{\pm} events 
    as a function of the center-of-mass photon energy, normalized to the value for $E^\star_\gamma=0$.
    The $E^\star_\gamma$ spectrum used in the MC is shown in the inset.\label{fig:kl3g}}}
Since kinematic closure of the event (\emiss=0) is not required for the 
identification of signal decays,
events with radiation are included (\Dkltreg{\pm}).
However, the acceptance for such events depends on the photon energy.
The MC simulation includes final-state radiation~\cite{MCgene_Gatti},
and allows the photon-inclusive reconstruction efficiency to be
determined for \ketreg{} and \kmutreg{} events.
The inclusion of radiation effects in the \kltre{} final state 
modifies the shape of the \mdue\ distributions used as inputs to
the fit used to count signal events, improving the fit quality.
The MC efficiency for \ketre{} events as a function of the photon energy in the \kao{} rest frame, $E^\star_\gamma$,
is shown in \Fig\ref{fig:kl3g}.
The fraction of events with $E^\star_\gamma$ greater than a 
reference energy $E_{{\rm ref}}$ is important only for \ketre{} decays.
For $E_{{\rm ref}}=20$~\MeV, for example, this fraction is 2.8\% for \ketre{}
decays and 0.1\% for \kmutre{} decays.

\section{Systematic uncertainties}
\label{sec:sys}

The systematic uncertainties in the \kltre{} \BR\ measurements are listed in \Ta\ref{tab:l3errosummary}.
All sources of systematic error have been 
evaluated for each tag sample and for each decay.
\TABLE{
  \begin{tabular}[c]{ccc}\hline
  Branching ratio         & \ketre{} & \kmutre{}\\ \hline 
  Source                  & \multicolumn{2}{c} {Fractional statistical error (10$^{-2}$)}\\ \hline
  Event counting          & 0.3      & 0.3      \\ 
  Tag bias                & 0.1      & 0.1      \\
  \effi{Sel}{}            & 0.7      & 0.8      \\ 
  \effi{FV}{}             & 0.2      & 0.4      \\ \hline
  {\bf Total statistical} &{\bf 0.8} &{\bf 0.9} \\ \hline
  Source                  & \multicolumn{2}{c} {Fractional systematic error (10$^{-2}$)} \\ \hline
  Event counting          & 0.2      & 0.1      \\ 
  Tag bias                & 0.3      & 0.3      \\
  \effi{Sel}{}            & 0.6      & 0.5      \\ 
  Stability               & 0.2      & 0.5      \\
  \effi{FV}{}             & 0.2      & 0.2      \\ \hline
  {\bf Total systematic}  &{\bf 0.7} &{\bf 0.8} \\ \hline\hline
  {\bf Total}             &{\bf 1.1} &{\bf 1.2} \\ \hline
  \end{tabular}
  \caption{Summary of contributions to the uncertainties for BR measurements.}
  \label{tab:l3errosummary}}

To evaluate the corrections to the decay-chain, photon-cluster, and 
lepton-cluster efficiencies 
(\rcorr{dc}, $ r_{\pai{0}}$, and \rcorr{lept}, respectively),
one or more control samples have been selected, and the efficiencies have 
been measured using these samples.
Each efficiency is evaluated in bins of a suitable set of physical variables, 
and the ratio of data and MC efficiencies is inserted into the MC.
The corrected signal efficiency is obtained by averaging over MC event distributions.

%%{\bf 1. Decay chain efficiency.} 
The decay-chain reconstruction efficiency correction 
has been measured using a control sample
of $\kao{\pm}\to\pai{0}X$ events identified in the \ntkmudue\ tag sample. 
The correction has been parameterized 
as a function of 
the kaon polar angle (\tk),
the decay vertex position (\rxy),
and the lepton momentum (\plab). 
The average data/MC correction is \rcorr{dc}\ab0.87 
and is mainly due to data-MC differences in the reconstruction efficiency 
for the kaon track.
The large energy loss of the \kao{\pm} 
in the DC gas\footnote{The DC gas is 90\% helium and 10\% isobutane.}
is underestimated in the MC, which results in a higher efficiency
for kaon track reconstruction.
To check the reliability of the correction applied, 
the events in each tag sample have been divided into equally populated and
statistically 
independent subsamples with \rxy\ less than and greater than 80~\cm,
and with $|\tk-90^\circ|$ less than and greater than $13^\circ$.
We find that the correction is larger ($1-\rcorr{dc}\sim0.20$)
for the samples with \rxy$<$80~\cm\ or $|\tk-90^\circ|>13^\circ$ 
than it is for the complementary samples with \rxy$>$80~\cm\ or 
$|\tk-90^\circ|<13^\circ$  ($1-\rcorr{dc}\sim0.05$).
The branching ratios measured using the full sample and
in the two subsamples for each decay and tag type coincide within the errors.
The systematic error has been taken to be half of the difference
between the BRs measured for the two subsamples in \rxy.
The fractional uncertainty in the BR measurements from \rcorr{dc} is about 0.54\% for \ketre{} 
and about 0.44\% for \kmutre{}, respectively.

%%{\bf 2. Lepton cluster efficiency.} 
The correction for the single lepton-cluster association efficiency is 
$1-\rcorr{\lep}\ab-0.004$ for electrons and $1-\rcorr{\lep}\ab0.02$ for muons.
Both lepton-cluster and decay-chain corrections strongly depend on \plab.
We have checked the stability of the BR measurements when 
\plab\ is additionally required to be greater than 50, 70, and 90\MeV.
We obtain a systematic error of about 0.2\% for \BR(\kmutre{}),
and a negligible error for \BR(\ketre{}).

%%{\bf 3. Photon cluster efficiency.}
The $r_{\pai{0}}$ correction takes into account differences between
data and MC in the  
cluster reconstruction efficiency for low-energy photons.
The single-photon detection efficiencies are evaluated from control samples of \kpidue{\pm} events, which are selected
using DC information only. 
A photon from \pai{0} decay is identified by requiring that its energy 
and time of flight be consistent with \kpidue{\pm} kinematics.
This provides a good estimate of the momentum of the second photon.
The efficiency is obtained as the probability for the second photon to be
found in a cone with opening angle $\cos{\tope}=0.7$ about the 
expected direction.
Each photon in the MC is weighted with the data/MC ratio of single-photon 
detection efficiencies evaluated as a function
of photon energy.
We have studied the effects on the correction $r_{\pai{0}}$ 
when the value of the opening-angle cut \tope\ is varied between 
 $\cos{\tope}=0.6$ and $\cos{\tope}=0.9$
and when the cut on the miminum energy for photon clusters is varied 
between 10 and 40~\MeV. 
We obtain a contribution to the uncertainties on the BRs of
about 0.2\% from photon-cluster systematics.
 
%%{\bf 4. Fit.}
In order to evaluate the systematic error
associated with the fit procedure,
we have performed various studies
using the \ntkmudue\ control sample (see \Se\ref{sec:tag}).
First, we use MC distributions in \mdue\ for \ketre{} and \kmutre{}
taken from the \ntkmudue\ sample without applying
the background-rejection cuts, which can in principle modify the shape
of the distributions.
We perform an additional check using \ketre{} and \kmutre{} fit shapes obtained directly from data.
Electron and muon cluster can be distinguished by exploiting the EMC granularity.
Cuts on the profile in depth of the energy deposited in the lepton cluster allow the selection of \ketre{}
(energy mainly deposited in the first EMC plane) or \kmutre{} (muons behave like minimum ionizing particles in the 
first plane while they deposit a sizeable fraction of their kinetic energy from the third plane onward) events.
This allows to obtain \ketre{} and \kmutre{} fit shapes directly from data.
We have tested the stability of the results when using these shapes.
Finally, we have checked that the results are stable against changes in
the histogram binning and fit range.
From these studies,
we estimate the fractional systematic uncertainty associated with the
fit procedure to range from 0.1\% to 0.4\%, depending on the decay mode and
tag type.

%%{\bf 5. Acceptance.}
\effi{FV}{} has been computed using the MC. 
For \kao{-} decays, \effi{FV}{} is corrected for losses due to 
nuclear interactions.
In this case, a contribution to the systematic error is evaluated from
the difference between the corrections measured for MC and data.
Actually, a suitable selection of \kao{\pm}$\to$\pai{0}$X$ events provides a 
sample containing \kao{\pm} interacting on the beam pipe and on the inner DC wall,
and therefore allows comparison of the effects of nuclear interaction in data and MC.
We obtain a fractional contribution of 0.37\% for \kmudue{+} tagged events
and 0.69\% for \kpidue{+} tagged events.

%%{\bf 6. Tag bias.}
The tag bias \atb\ includes the effect of the \filfo\ and cosmic-ray veto (\cveto) filters.
The \filfo\ correction has been measured for each tag sample separately. It is about 0.1\% 
for \kpidue{}-tagged events and about 1.5\% for \kmudue{}-tagged events.
The systematic error has been conservatively taken to be equal to the correction itself.
For the \cveto, the measured correction ranges from 0.04\% to 0.09\%, depending on the tag sample;
we assign a systematic error equal to half the value of the correction itself.
Finally, since \kao{-} losses to nuclear interactions contribute to the 
value of \atb, we assign an additional fractional error of \ab0.1\% 
for the \kmudue{+}- and \kpidue{+}-tagged samples.

%%{\bf 7. Stability.}
Last, we use the MC to check the stability of the results with respect to 
variations of each of the cuts used to increase the purity of the 
\kltre{\pm} samples.
Moving the \mtrenta\ cut from 50 to 70~\MeV\ changes the \ketre{} efficiency
from \ab0.89 to \ab0.77 and the \kmutre{} efficiency from \ab0.87 to \ab0.70,
while inducing variations of \ab0.1\% and \ab0.4\% in the resulting 
\ketre{} and \kmutre{} BRs, respectively.
We have performed similar studies for the \pstarp\ and \emiss\ cuts, 
giving a total contribution to the fractional systematic error of 
\ab0.17\% for \ketre{} and \ab0.49\% for \kmutre{}.

\section{Results}
\label{sec:result}

The four determinations of the \ketre{} and \kmutre{} branching ratios are 
listed in \Ta\ref{tab:brl3final}.
The fractional uncertainties range from 1.5\% to 2.1\% for \BR(\ketre{})
and from 1.5\% to 2.7\% for \BR(\kmutre{}).
\TABLE{
  \begin{tabular}{ccccc}\hline
    Tagging decay &\kmudue{+}    & \kpidue{+}     & \kmudue{-}   & \kpidue{-}  \\ \hline \hline
   \bketre{}      & 0.04953 (74) &  0.04930 (103) & 0.04968 (76) & 0.05024 (102) \\ \hline
   \bkmutre{}     & 0.03217 (63) &  0.03223 (87)  & 0.03233 (49) & 0.03275 (86)  \\ \hline
  \end{tabular}
  \caption{Final results for \bketre{} and \bkmutre{} measured with each tag sample.}
  \label{tab:brl3final}}
Averaging the results for each charge state, we obtain:
\begin{align*}
\BR(\ketre{-}) & = (4.946 \pm 0.053_{{\rm stat}} \pm 0.038_{{\rm syst}})\times10^{-2} \\
\BR(\ketre{+}) & = (4.985 \pm 0.054_{{\rm stat}} \pm 0.037_{{\rm syst}})\times10^{-2} {\mbox ,}
\end{align*}
and 
\begin{align*}
\BR(\kmutre{-}) & = (3.219 \pm 0.047_{{\rm stat}} \pm 0.027_{{\rm syst}})\times10^{-2} \\
\BR(\kmutre{+}) & = (3.241 \pm 0.037_{{\rm stat}} \pm 0.026_{{\rm syst}})\times10^{-2} {\mbox .}
\end{align*}
The \cdue\ between the measurements of \ketre{-} and \ketre{+} is 0.17/1 
(probability \ab0.68);
for  \kmutre{-} and \kmutre{+} it is 0.12/1 (probability \ab0.73).
The final averages are:
\begin{align*}
\BR(\ketre{ })  & = (4.965 \pm 0.038_{{\rm stat}} \pm 0.037_{{\rm syst}})\times10^{-2} \\
\BR(\kmutre{ }) & = (3.233 \pm 0.029_{{\rm stat}} \pm 0.026_{{\rm syst}})\times10^{-2}  {\mbox .}
\end{align*}
The \cdue\ for the four independent measurements for each tag type is 1.62/3
for \ketre{} decays (probability \ab0.65) and
and 1.07/3 for \kmutre{} decays (probability \ab0.78).
Our final \BR\ results have a fractional uncertainty of 1.1\% for \ketre{} 
and 1.2\% for \kmutre{} decays;
all contributions to the error are summarized in \Ta\ref{tab:l3errosummary}.
The dominant contribution to the total error is from the statistics 
used to estimate the correction to the \eff{\ltre} efficiency.

All of the averages and \cdue\ values quoted above are calculated 
with all correlations between measurements taken into account.
While the correlation between the numbers of \ketre{} and \kmutre{}
events induced by the fit procedure is low (about 1\%),
a significant correlation arises from the corrections to the tag bias, 
which are equal for the two channels,
as well as from the
data/MC corrections for the tracking and the clustering efficiencies, and 
finally, from the selection cuts.
Excluding the contribution from the uncertainty in
the value of the \kao{\pm} lifetime, the total error matrix for the 
final measurements of \bketre{} and \bkmutre{} is
\[ \left( \begin{array}{cccc}
0.2780 & & & 0.1268 \\ 
0.1268 & & & 0.1510
\end{array} \right) \times 10^{-6} {\mbox ,}\]
corresponding to a correlation coefficient between the errors on 
\bketre{} and \bkmutre{} of 62.7\%.

With this correlation taken into account, we evaluate
the ratio $\rme{}=\rappo{}$ from our results for \bketre{} and \bkmutre{}.
We obtain $\rme{}=0.6511\pm0.0064$.
This value has a fractional error of about 1.0\% and is in $1.5\sigma$
agreement with the theoretical
prediction, \rme{SM}=0.6646(61)~\cite{FlaviaNet06}.

Using the PDG value for the \kao{\pm} lifetime~\cite{PDG06} and
the KLOE values for the semileptonic form factors~\cite{KLOEklffmu},
we obtain
\begin{align*}
\Vusfo & = 0.2148 \pm 0.0013 \;{\rm ~from~\bketre{}~and} \\
\Vusfo & = 0.2129 \pm 0.0015 \;{\rm ~from~\bkmutre{}}.
\end{align*}
The average is \Vusfo\ = 0.2141$\pm$0.0013,
including the correlations between the \BR\ measurements and the use of the same lifetime value for both decays.
Using \fo = 0.961(8) from~\cite{LR84}, $|\Vus|$ is 0.2223(23).
With $|\Vud| = 0.97418(26)$~\cite{TH07}, we find 
$1 - |V_{ud}|^2 - |V_{us}|^2  = 0.0016(11)$,
so that the first-row test of the unitarity of the CKM matrix is 
satisfied at the level of $1.4\sigma$.

\section*{Acknowledgments} %New from ack0102_20071123.txt

We thank the \DAF\ team for their efforts in maintaining low background running 
conditions and their collaboration during all data-taking. 
We want to thank our technical staff: G.F.~Fortugno and F.~Sborzacchi for their dedicated work to ensure an
efficient operation of the KLOE Computing Center; 
M.~Anelli for his continuous support to the gas system and the safety of the detector; 
A.~Balla, M.~Gatta, G.~Corradi, and G.~Papalino for the maintenance of the electronics;
M.~Santoni, G.~Paoluzzi, and R.~Rosellini for the general support to the detector; 
C.~Piscitelli for his help during major maintenance periods.
This work was supported in part
by EURODAPHNE, contract FMRX-CT98-0169; 
by the German Federal Ministry of Education and Research (BMBF) contract 06-KA-957; 
by the German Research Foundation (DFG), 'Emmy Noether Programme', contracts DE839/1-4;
by INTAS, contracts 96-624, 99-37; 
and by the EU Integrated Infrastructure Initiative HadronPhysics Project under contract number RII3-CT-2004-506078.
\bibliographystyle{elsart-num}
\bibliography{biblio}
\end{document}